\documentclass{aa}  

\usepackage{fancyhdr}
\usepackage{graphicx}
\usepackage{txfonts}
\usepackage{lipsum}
\usepackage{subcaption}         
\usepackage{lscape}             
\usepackage{placeins}           

\usepackage{natbib}
\usepackage{hyperref}

\usepackage{listings}
\usepackage{xcolor}

\definecolor{aol-bg}{HTML}{FAFAFA}
\definecolor{aol-fg}{HTML}{383A42}
\definecolor{aol-comment}{HTML}{A0A1A7}
\definecolor{aol-red}{HTML}{E45649}
\definecolor{aol-orange}{HTML}{D19A66}
\definecolor{aol-yellow}{HTML}{C18401}
\definecolor{aol-green}{HTML}{50A14F}
\definecolor{aol-cyan}{HTML}{0184BC}
\definecolor{aol-blue}{HTML}{4078F2}
\definecolor{aol-purple}{HTML}{A626A4}

\lstdefinestyle{atomonelight}{
    backgroundcolor=\color{white},
    basicstyle=\ttfamily\small\color{aol-fg},
    keywordstyle=\color{aol-purple}\bfseries,
    stringstyle=\color{aol-green},
    commentstyle=\color{aol-comment}\itshape,
    identifierstyle=\color{aol-fg},
    emphstyle=\color{aol-red},
    morekeywords={self},
    numberstyle=\tiny\color{aol-comment},
    stepnumber=1,
    numbersep=10pt,
    showspaces=false,
    showstringspaces=false,
    showtabs=false,
    frame=single,
    rulecolor=\color{aol-comment},
    tabsize=4,
    breaklines=true,
    breakatwhitespace=true,
    captionpos=b,
    keepspaces=true
}

\begin{document}

   \title{Enabling fundamental understanding of Nature \\ with novel binning methods for 2D histograms}


%
%
%

   \author{I. Vaiman\inst{1, 2}\corrauth{igor.vaiman@gssi.it}}

   \institute{
       Gran Sasso Science Institute (GSSI), Viale Francesco Crispi 7, 67100 L’Aquila, Italy
   \and 
      INFN-Laboratori Nazionali del Gran Sasso (LNGS), via G. Acitelli 22, 67100 Assergi (AQ), Italy
   }

   \date{Received April 1, 2026}

  \abstract
    {Visualization of 2D distributions is an essential task, commonly done with a 2D histogram.
    The histogram is built by subdividing the sample space into regions and color-coding the number of
    samples in each region.}
    {We aim to solve long-standing problems with common 2D histogram methods: lack of thematic,
    visual, and conceptual unity with underlying data, and general stagnation in the field.} 
    {We develop a new method for plotting 2D histograms with arbitrary bin shapes,
    including aperiodic tilings and geographic maps. We apply the method to several
    common plot types from the literature.}
    {We find our method performs best across all tasks, solving the problems
    and propelling the scientific progress forward.}
    {}

   \maketitle

    \nolinenumbers

\section{Introduction}

Visualizing two-dimensional distribution samples is essential for reporting any
modern scientific result. The main tool for doing that is the 2D histogram,
constructed by dividing the sample space into regions (bins) and using color to
show the number or density of samples in a given bin. Yet, despite the immense
importance of 2D histograms, for a long time, data visualization practitioners have
faced debilitating stagnation. Most tools (e.g., Matplotlib \citep{Hunter:2007},
ROOT \citep{ROOT_NIMA_1997}) limit them to simple rectangular or,
at best, hexagonal bins. Some domain-specific binning methods, such
as HEALPix \citep{healpix}, exist, but their applications are naturally limited.

To solve this long-standing issue in scientific and broader data analysis
communities, we propose a novel method: \textbf{F}undamental \textbf{U}nderstanding
of \textbf{N}ature \textbf{Bin}ning, abbreviated \verb|funbin|.
It allows one to produce 2D histograms using a wide variety of bin shapes, including
but not limited to periodic and aperiodic plane tilings, Voronoi diagrams,
outlines of geographical areas, and others. The paper is structured as follows:
in Section \ref{sec:foundation} we outline the conceptual foundations of \verb|funbin|,
drawing on modern and ancient philosophical and mystical traditions; in
Section \ref{sec:method} we describe a reference \verb|funbin| implementation as
a \verb|Matplotlib|-compatible Python package; in Section \ref{sec:examples} we
illustrate the depth and breadth of scientific breakthroughs to be brought about
by \verb|funbin| with a few examples from astro- and particle physics;
finally, we conclude in Section \ref{sec:discussion}.

\section{Conceptual foundation}
\label{sec:foundation}

Hermes Trismegistus is an ancient mystical and philosophical figure who features characteristics
of both Hermes and Thoth. The former is a Greek god of travelers, merchants, and orators. The latter
is an Egyptian god of, among others, wisdom and learning. It is therefore not surprising that
a modern scientist, struggling for knowledge, but also precariously searching
to sell this knowledge, can find inspiration in the writings of Hermes Trismegistus.

To outline the conceptual foundation of \verb|funbin|, we draw on one of the most famous texts by the
"thrice great", the Emerald Tablet \citep{EmeraldTablet}. In turn, one of the most famous quotes
from the text says (the translation from \citet{HOLMYARD1923}):

\begin{quote}
    That which is above is from that which is below, and that which is below is from
    that which is above, working the miracles of one [thing]. As all things were from One.
\end{quote}

In medieval and Renaissance times, this principle was adopted to state the similarity
between the Microcosm and Macrocosm, for example, in \cite{Paracelsus}. It is now our turn,
not dissimilar to Paracelsus, to base our practices on near-inscrutable writing from centuries
long gone. In the context of this work, the Macrocosm represents the objects under study in the
world. The Microcosm is, in turn, the interior region of a scientific work, most importantly, the plots. Therefore, the adaptation of Hermeto-Paracelsian insight to scientific practice
is to find visual, conceptual, and thematic parallels between objects under study and the way
one visualizes them.

If we examine the types of distribution one usually aims to visualize, we readily
find almost none of them to look like squares. Nature rarely produces $C_4$-symmetric shapes,
instead favoring fuzzy blobs. Therefore, the widespread use of square or rectangular bins
comes into pronounced conflict with the vast majority of the Macrocosm. One might argue that
hexagonal bins improve on the situation by being closer to a circular shape (continuous circular
symmetry) while still tiling the plane. However, in our view, replacing the unnatural $C_4$ symmetry
with a slightly less unnatural $C_6$ is merely a band-aid solution.

Moreover, both rectangular and hexagonal bins tile the plane in a periodic way. More precisely,
both tilings feature global translational symmetry, that is, some translations
map the tiling onto itself. This fact alone sets scientific research up for failure: just like
the grid one uses, one is forced to indefinitely repeat the same pattern.

Finally, even if one manages to escape the trap of endless repetition and the pitfalls
of squaring non-squares, the ubiquity of conventional binnings leads to the lack of conceptual
resonance between the Micro- and Macrocosm. We might imagine rare cases where this is not true:
a lattice QCD researcher using square binning to mirror the spacetime discretization, or a
melittologist binning their data in hexagonal bins to evoke an image of the honeycomb. However,
in most cases, the conventional binnings do not evoke any image in the Macrocosm of the work.

\section{Method}
\label{sec:method}

We have implemented the principles of \verb|funbin| in an eponymous Python package\footnote{
\url{https://github.com/nj-vs-vh/funbin}
}. The package comprises two main parts: the main plotting function, also called \verb|funbin|,
and a suite of algorithms to generate polygon
sets for various binnings. Below is a self-contained example script:

\begin{lstlisting}[language=Python]
import numpy as np
from numpy.random import normal, random
from matplotlib import pyplot as plt
from funbin import funbin
from funbin.einstein import aperiodic_monotile

# mixing 80/20 between two offset Normal modes
n = 50000
samples = np.where(
    random(n) > 0.2,
    normal(0, 1.0, (2, n)),
    normal([[2.0], [2.0]], 0.5, (2, n))
)

# tiling is simply a list of Polygon objects
tiling = aperiodic_monotile(bins=(40, 40))

# API largely mirrors ax.hexbin
fig, ax = plt.subplots() 
funbin(ax, *samples, tiling=tiling)
ax.set_aspect("equal")
fig.savefig("funbin-example.png")
\end{lstlisting}
The function \verb|funbin| admits an arbitrary set of polygons to be used as bins. The
The algorithm is straightforward:
\begin{enumerate}
    \item Translate and stretch polygons to match the data bounding box.
    \item For every data point, find a polygon it belongs to. This is done with the help of
    an auxiliary "spatial index" that rejects far-away polygons without running the
    point-in-polygon algorithm.
    \item Compute sample weight sum for each polygon. For an unweighted histogram, all weights
    are $1$, so the weight sum is a simple count.
    \item Divide each polygon's weight sum by its area to estimate sample weight density
    within its boundaries.
    \item Plot polygons with their colors defined by the sample weight density and a colormap.
\end{enumerate}

Algorithms to generate polygon sets include:
\begin{itemize}
    \item Penrose tilings P1, P2, P3 \citep{Penrose1979}
    \item Aperiodic monotile \citep{Smith2024}
    \item Voronoi diagrams \citep{Voronoi1908} of a random or user-specified set of points
    \item Map data imported from GeoJSON format \citep{rfc7946}.
\end{itemize}

\section{Examples}
\label{sec:examples}

In this section, we apply \verb|funbin| to several example problems and highlight
how the chosen binning (Microcosm) enhance one's fundamental understanding of the
problem at hand and Nature as a whole (Macrocosm). We first focus on two aperiodic
tilings: Penrose tiling of three types and a recently discovered aperiodic monotile
with two variants on the tile shape. Their aperiodicity itself automatically elevates
the scientific value of the plotted data, since it breaks the malignant repetitiveness
associated with conventional periodic tilings, as described in Section \ref{sec:foundation}.
For each tiling, we find a particular dataset that is especially concordant with its shape.
Finally, we explicitly explore the Microcosm-Macrososm resonance by using Earth's
geographical data to map out stellar mass distribution in the Universe.

\subsection{Penrose tilings}
\label{subsec:penrose}

\begin{figure}[t]
    \centering
    \includegraphics[width=\columnwidth]{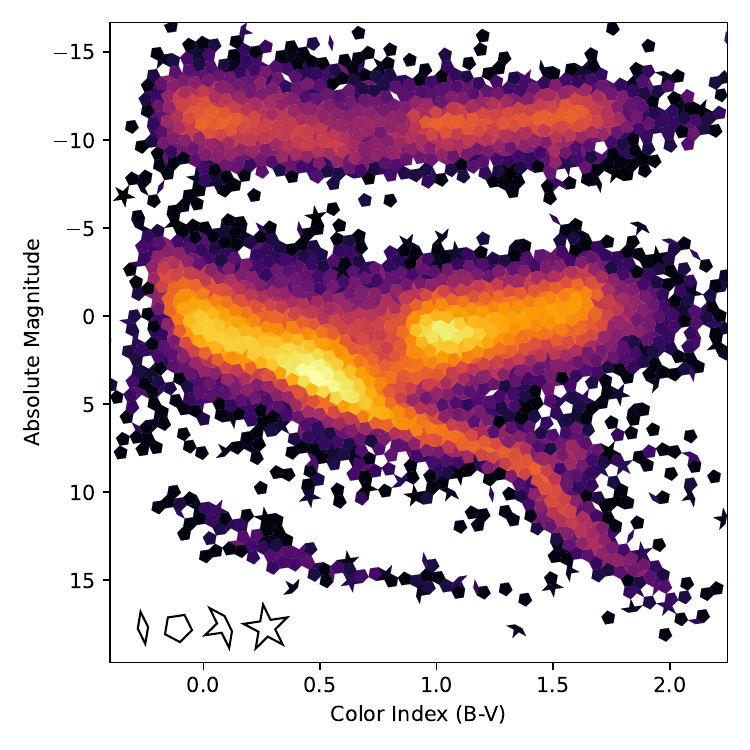}
    {\caption{Hertzsprung-Russell diagram binned with Penrose P1 tiling.
    Data from HYG (Hipparcos, Yale, Gliese) Stellar database is used.
    Bottom left corner shows tile shapes.}
    \label{fig:hg}
    }
\end{figure}

Penrose tilings are a set of aperiodic tilings described in \citet{Penrose1979}.
The first of them chronologically, P1 tiling consists of five shapes: three kinds of
pentagons (they look the same but have different matching rules, so are considered
different tiles), a star, a diamond, and a boat. The imagery is hard to miss:
aside from a literal star, the boat is essentially a broken star, perhaps disrupted
by tidal forces or collision; the pentagon is nothing but a star with no rays coming out
from it, an extinct star, and the three similarly-looking pentagons are three different
endpoints of stellar evolution: black hole, neutron star, white dwarf; finally, the
diamond can be interpreted as referring to the crystalline structure at the cores of
carbon-rich white dwarfs. Driven by these considerations, in Figure \ref{fig:hg}
we plot data on observed star color and magnitude from the HYG database \citep{hyg_database}
as a binned Hertzsprung-Russell diagram, using Penrose P1 tiling as bins.

\begin{figure}[t]
    \centering
    \includegraphics[width=\columnwidth]{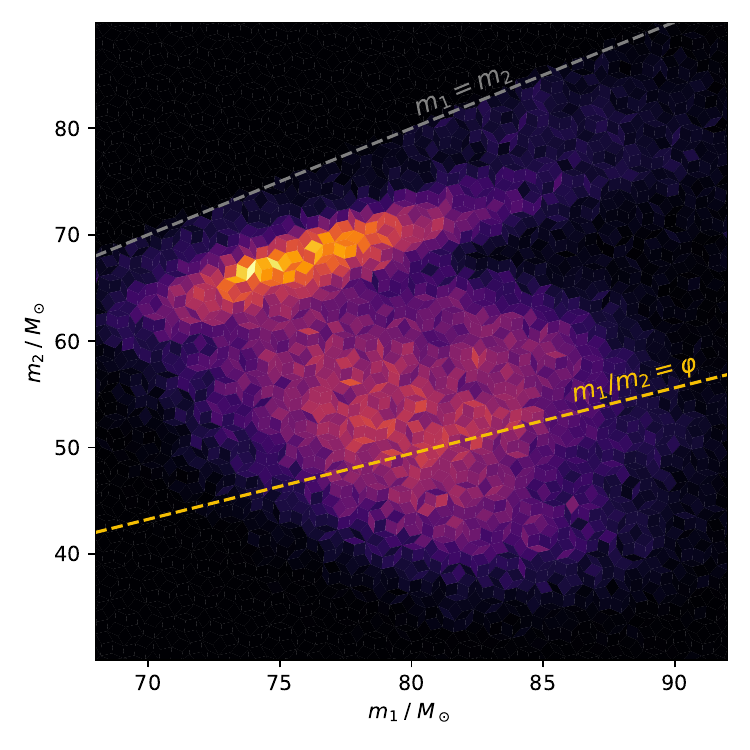}
    \caption{
        Posterior distribution of black hole masses from
        \texttt{GW191109\_010717} binned with Penrose P2 tiling.
    \label{fig:bbh}
    }
\end{figure}

The most widely known Penrose tiling is P3, consisting of two types of rhombus, 
sometimes dubbed "fat" and "thin". Another situation where two similar but somewhat
differently-sized objects come together to rather spectacular results is a black hole
merger. It is not surprising then that Roger Penrose himself has contributed greatly
to the study of black holes. In the last ten years, great progress has been achieved
in observing binary black hole mergers through gravitational wave interferometry \citep{Abbott2016}.
Gravitational wave astronomy has also been one of the leading fields in applying 
Bayesian statistics for event reconstruction, which in practice implies working
with multidimensional samples from the posterior distribution of interest. All this
makes GW astronomy another field to be readily revolutionized by \verb|funbin|.

To illustrate this, in Figure \ref{fig:bbh} we plot a posterior distribution
of two black holes' masses (marginalized over all other parameters) from the event
\verb|GW191109_010717| published by the LVK collaboration as part of the GWTC-3
parameter estimation data release \citep{LVK_Open_Data}. We notice that the two
types of rhombus used in the tiling, combined with gradient shading from the posterior
distribution, create structures resembling projections of 3D cubes. This emergent
property highlights the power of \verb|funbin| to show higher-dimensional
structures in the data, even when technically it is limited to two. Moreover,
we notice that a secondary mode of the posterior features black holes with a mass
ratio close to the golden ratio $\varphi \approx 1.618$, shown in a golden dashed line.
Even if this is not remarkable on its own, when paired with the fact that the golden ratio also 
gives the ratio between areas of fat and thin rhombi, this clearly shows the ability of
\verb|funbin| to highlight otherwise obscure mathematical relations in the data it
visualizes.

\begin{figure}[t]
    \centering
    \includegraphics[width=\columnwidth]{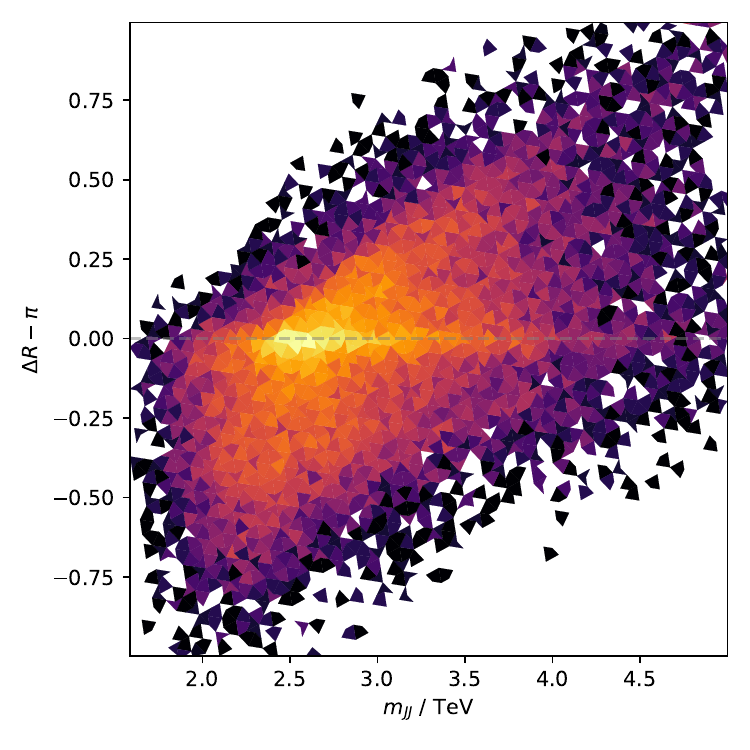}
    {\caption{
        Total invariant mass of two heaviest jets $m_{JJ}$ versus the $\eta$-$\phi$ cylinder
        distance between them from the LHCO2020 BlackBox1 dataset, binned with Penrose P3 tiling.
    }
    \label{fig:lhco}
    }
\end{figure}

Finally, the Penrose P2 tiling uses two types of tiles, sometimes dubbed "darts"
and "kites". Both darts and kites are designed to fly, and darts are specifically
designed to precisely hit a target when thrown from some distance. Perhaps the most
impressive human-made dart-throwing device is the Large Hadron Collider with multi-TeV
proton darts. Also, the dart and kite shapes can be reinterpreted as symbolic representations of
particle jets inside of the collider. Larger intensity in the central region of the jet
gives the kite-shaped jet profile, while larger intensity in the shell --- the dart-like
one. For these reasons, in Figure \ref{fig:lhco} we plot LHC simulation data
from LHC Olympics 2020 \verb|BlackBox1| dataset \citep{LHCO2020} with Penrose P2 binning. We use
\verb|FastJet| \citep{Cacciari2012} to analyze simulated jets and plot the total invariant mass
of the two heaviest jets against the $\eta$-$\phi$ cylinder distance between them.
The central purpose of the LHCO dataset is to provide a benchmark for anomaly detection
algorithms. However, we note that by using a non-standard binning and a potentially buggy
implementation \verb|funbin| allows one to instead inject new anomalies into the data,
opening exciting new avenues for high-energy physics.

\subsection{Aperiodic monotile}

As explained in Section \ref{subsec:penrose}, aperiodicity itself is a highly desirable
property for binnings used in scientific plots. Until recently, only two-tile
aperiodic tiling, such as P2 and P3, have been known. But in 2023, a single shape that can
tile the whole plane with no repeats, an aperiodic monotile, was discovered \citep{Smith2024}.
It is also called einstein, from "ein Stein", meaning "one stone" in German. The addition of
singularity to aperiodicity has profound implications for the Hermeto-Paracelsian
reasoning: the tiling is built on a single principle (shape), yet it never repeats itself. Thus,
it provides a geometrical foundation for the scientific method, which is also a singular 
principle, yet it is supposed to facilitate an endless non-repeating procession of 
scientific activity.

\begin{figure}[t]
    \centering
    \includegraphics[width=\columnwidth]{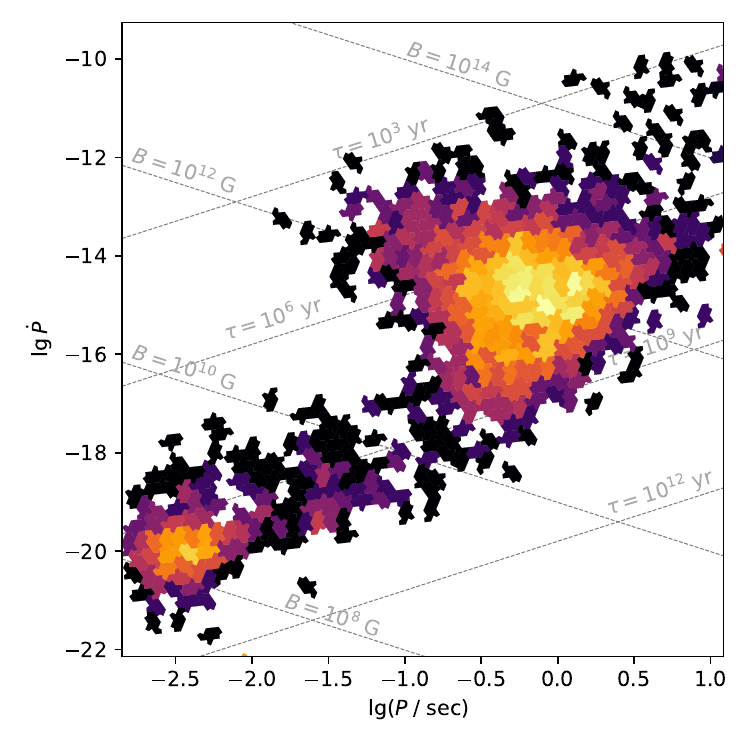}
    {\caption{
        $P$ - $\dot P$ diagram for pulsars from ANTF Catalog \citep{Manchester2005} binned with aperiodic
        monotile "turtle".
        Lines of negative slope show constant characteristic lifetime $\tau \equiv \frac{P}{2 \dot P}$
        and lines of positive slope show constant characteristic magnetic field
        $B \equiv \sqrt{\frac{P}{\text{sec}} \dot P} \; 3.2 \times 10^{19}~\text{G}$
    }
    \label{fig:pulsars}
    }
    
\end{figure}

Inspired by meditations on the nature of (a)periodicity, we turn our
attention to pulsars. Pulsars are highly magnetized and fast-rotating neutron stars, and they
constitute probably the most extreme case of periodic behavior in the Universe. At the same time,
they generally show measurable change in their period over time, slowly radiating away the rotational
energy, therefore their true nature is aperiodic. For our visualization, we retrieve a sample of
2798 pulsars from ATNF Pulsar Catalog \citep{Manchester2005} \footnote{
\url{https://www.atnf.csiro.au/research/pulsar/psrcat/}
} and reproduce the classic $P$ - $\dot P$ diagram in Figure \ref{fig:pulsars}.
For binnning, we use the "turtle" (Tile$ \left( \sqrt{3}, 1 \right)$) shape from the aperiodic monotile
family. The name calls to mind the fable of the tortoise (turtle's relative) and the hare: like the
hare, the pulsar is very fast, but is recklessly slowing down and sooner or later (in
$\tau = 10^3$ to $10^6$ years) risks being overtaken even by a turtle. Aside from profound
scientific insights, \verb|funbin| is capable of giving life advice.

\begin{figure}[t]
    \centering
    \includegraphics[width=\columnwidth]{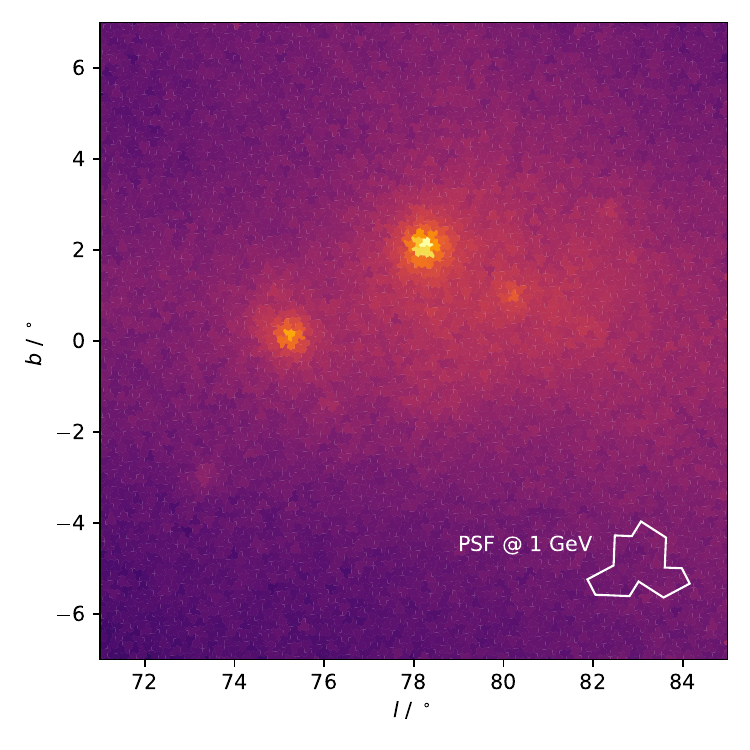}
    {\caption{
        Raw $\gamma$-ray counts detected by Fermi LAT in the sky area around PSR J2032+4127,
        binned with aperiodic monotile "hat".
    }
    \label{fig:fermi}
    }
    
\end{figure}

In another pulsar-related application, in Figure \ref{fig:fermi} we use a different aperiodic monotile 
from the same family, "hat" (Tile$ \left(1,  \sqrt{3} \right)$), as a pixel for $\gamma$-ray data
from Fermi LAT \citep{Atwood2009}. Namely, we have selected $0.2 - 10$ GeV photons in a $22^\circ$
radius around the direction of PSR J2032+4127 ($\text{RA} = 308.055, \text{Dec} = 41.4568$)
recorded during the year 2025. The data show slowly varying diffuse emission and three bright sources,
pulsars J2021+4026, J2021+3651, and J2032+4127. The bottom right corner shows the Fermi LAT point-spread function around $1$ GeV energy if it were shaped like the hat tile.

\subsection{World map}

\begin{figure*}[t!]
    \centering
    \includegraphics[width=\textwidth]{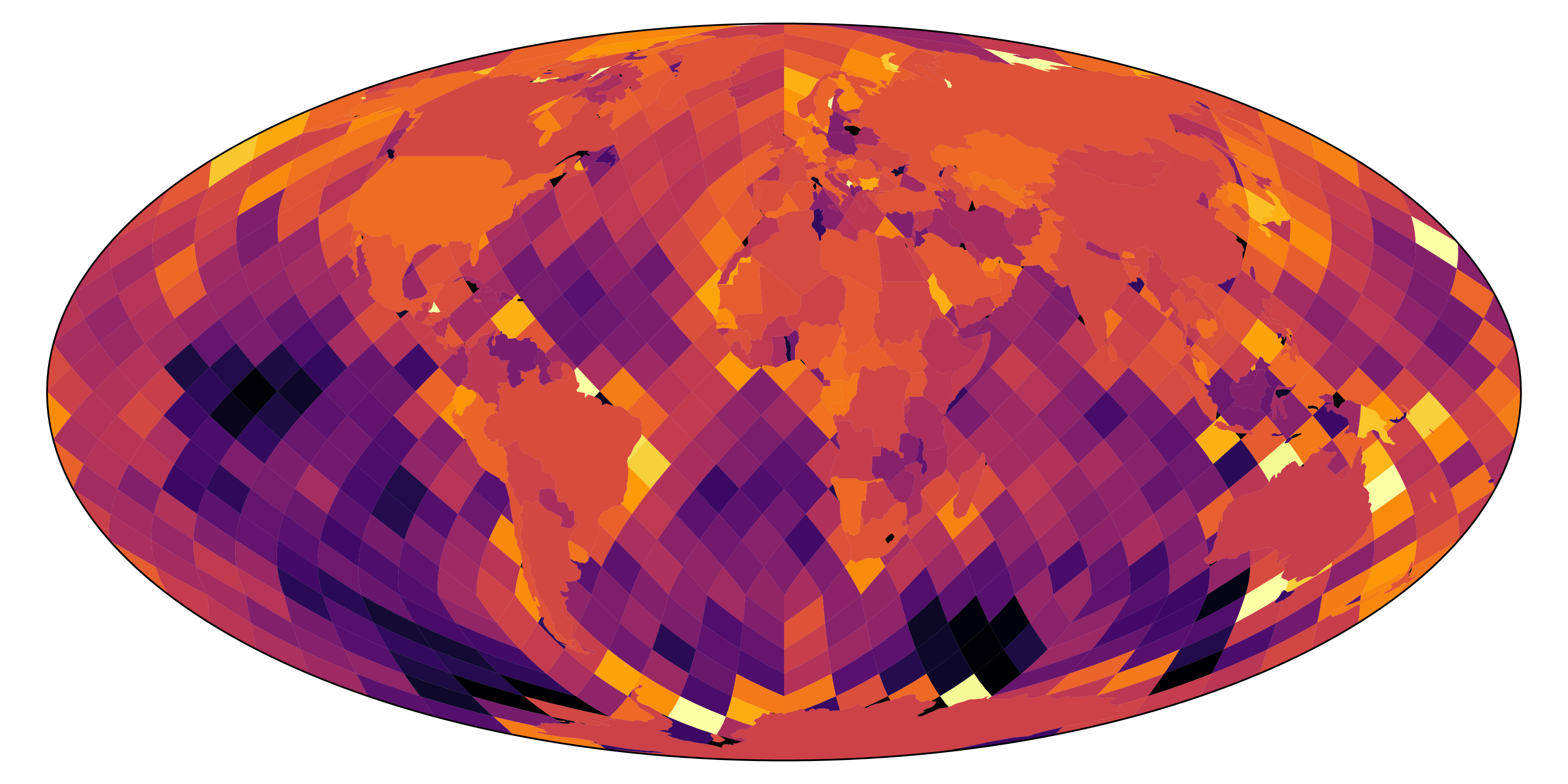}
    {
        \caption{
        Galaxies  in the local Universe ($d_L < 200$ Mpc) from \cite{Biteau2021}, weighted by their
        stellar mass and shown in Galactic coordinates, binned with Earth's political map and
        HEALPix subdivision of the oceans.
        }
    \label{fig:sky}
    }
\end{figure*}

In this section, we achieve an explicit Microcosm/Macrocosm convergence by bringing together
two spheres: Earth's political map provides binning for stellar mass distribution
across the local Universe within 200 Mpc from \cite{Biteau2021}. For the former, we use the Natural Earth
dataset \citep{naturalearth} to extract administrative boundaries of all countries at $110$ m
precision. Since Earth's oceans do not have administrative boundaries, they would constitute
a single giant bin, which leads to a poorly readable plot. Therefore,
we subdivide the ocean with the regular HEALPix bins ($n_\text{side} = 8$). For the data we extract
from \cite{Biteau2021} galaxies with $d_L < 200$ Mpc and weight them according to their estimated
$M_\star$, accounting for the mass incompleteness correction. We then juxtapose their Galactic longitude 
and latitude onto Earth's longitude and latitude and our combined countires and ocean pixels binning.
The resulting 2D histogram in the Mollweide projection is shown in Figure \ref{fig:sky}.

\verb|funbin| allows us to draw profound conceptual conclusions here. By juxtaposing the spheres
of Earth and the Universe, we essentially rediscover Kantian epistemology: the Earth’s map serves
as a region of transcendental, a frame that structures our experience of the Cosmos. Perhaps there is
already a post-Kantian, even Marxist twist here: the transcendental appears to be historically conditioned
and produced. This leads to a speculative connection of \verb|funbin| with politics.
For example, if one desires the plot to have greater angular resolution in the region of the Virgo
cluster, one is naturally drawn to the political goal of breaking a large pixel of modern-day Russia
down into smaller ones. One could also make the case for abolishing the Antarctic Treaty and
dividing Antarctica into countries to gain resolution in the South Galactic Pole region.

We conclude this brief exploration with another famous and relevant quote from
\cite{kant1999critique}:

\begin{quote}
Two things fill the mind with ever-increasing wonder and awe,
the more often and the more intensely the mind of thought is drawn to them:
The starry heavens above me and the moral law within me.
\end{quote}

It is regrettable, but not unexpected, that the role of the moral law in this analogy
is played by the borders of nation-states.

\section{Discussion}
\label{sec:discussion}

We have presented our solution to long-term problems in the field
of data visualization, namely concerning 2D histogram plots:
\textbf{F}undamental \textbf{U}nderstanding
of \textbf{N}ature \textbf{Bin}ning, or \verb|funbin|. We have
outlined multiple issues with existing bin shapes (rectangular and hexagonal),
namely, a lack of visual, thematic, and philosophical harmony between them and
the data they are supposed to represent. We draw on ancient, medieval, and
early modern mysticism and philosophy to create a Hermeto-Paracelsio-Kantian
framework, which we apply to several data visualization tasks from astro-
and particle physics. We find that \verb|funbin| solves a significant
portion of the problems (while also introducing interesting new ones),
thus constituting a significant step in the scientific progress.

\begin{acknowledgements}
      Part of this work was supported by the need to take a 14-hour flight.
      Artificial Intelligence tools were not used in its preparation,
      while Natural Intelligence was used sparingly.
\end{acknowledgements}

\bibliographystyle{abbrvnat} 
\bibliography{main}

\end{document}